\documentclass[12pt]{article}
\usepackage[dvips]{graphicx}

\begin{document}
\title{Quantum Magnetohydrodynamics}
\author{
F.~Haas\footnote{ferhaas@exatas.unisinos.br}\\
Universidade do Vale do Rio dos Sinos - UNISINOS \\
Unidade de Exatas e Tecnol\'ogicas \\
Av. Unisinos, 950\\
93022--000 S\~ao Leopoldo, RS, Brazil}
\maketitle
\begin{abstract}
\noindent The quantum hydrodynamic model for charged particle systems is extended to the cases of non zero magnetic fields. In this way, quantum corrections to magnetohydrodynamics are obtained starting from the quantum hydrodynamical model with magnetic fields. The quantum magnetohydrodynamics model is analyzed in the infinite conductivity limit. The conditions for equilibrium in ideal quantum magnetohydrodynamics are established. Translationally invariant exact equilibrium solutions are obtained in the case of the ideal quantum magnetohydrodynamic model.  
\end{abstract}

\hspace{.25cm} {{\it PACS numbers: 52.30.Cv, 52.55.-s, 05.60.Gg}}

\section{Introduction}

There has been an accrued interest on quantum plasmas, motivated by applications in ultra small electronic devices \cite{Markowich}, dense astrophysical plasmas \cite{Chabrier}-\cite{Jung} and laser plasmas \cite{Kremp}. Recent developments involves quantum corrections to Bernstein-Greene-Kruskal equilibria \cite{Luque}, quantum beam instabilities \cite{Anderson}-\cite{Haas3}, quantum ion-acoustic waves \cite{Haas4}, quantum corrections to the Zakharov equations \cite{Garcia, Haas5}, modifications on Debye screening for quantum plasmas with magnetic fields \cite{Shokri1}, quantum drift waves \cite{Shokri2}, quantum surface waves \cite{Shokri3}, quantum plasma echoes \cite{Manfredi1}, the expansion of a quantum electron gas into vacuum \cite{Mola} and the quantum Landau damping \cite{Suh}. In addition, quantum methods have been used for the treatment of classical plasma problems \cite{Fedele, Bertrand}. 

One possible approach to charged particle systems where quantum effects are relevant is furnished by quantum hydrodynamics models. In fact, hydrodynamic formulations have appeared in the early days of quantum mechanics \cite{Madelung}. More recently, the quantum hydrodynamics model for semiconductors has been introduced to handle questions like negative differential resistance as well as resonant tunneling phenomena in micro-electronic devices \cite{Gardner1}-\cite{Gardner3}. The derivation and application of the quantum hydrodynamics model for charged particle systems is the subject of a series of recent works \cite{Manfredi2}-\cite{Degond}. In classical plasmas physics, fluid models are ubiquitous, with their applications ranging from astrophysics to controlled nuclear fusion \cite{Nicholson, Bittencourt}. In particular, magnetohydrodynamics provides one of the most useful fluid models, focusing on the global properties of the plasma. The purpose of this work is to obtain a quantum counterpart of magnetohydrodynamics, starting from the quantum hydrodynamics model for charged particle systems. This provides another place to study the way quantum physics can modify classical plasma physics. However, it should be noted that the quantum hydrodynamic model for charged particle systems was build for non magnetized systems only. To obtain a quantum modified magnetohydrodynamics, this work also offer the appropriated extension of the quantum hydrodynamics model to the cases of non zero magnetic field.  

The paper is organized as follows. In Section 2, the equations of quantum hydrodynamics are obtained, now allowing for the presence of magnetic fields. The approach for this is based on a Wigner equation with non zero vector potentials. Defining macroscopic quantities like charge density and current through moments of the Wigner function, we arrive at the desired quantum fluid model. In Section 3, we repeat the well known steps for the derivation of magnetohydrodynamics, now including the quantum corrections present in the quantum hydrodynamic model. This produces a quantum magnetohydrodynamics set of equations. In Section 4, a simplified set of quantum magnetohydrodynamics is derived, yielding a quantum version of the generalized Ohm's law. In addition, the infinity conductivity case is shown to imply an ideal quantum magnetohydrodynamic model. In this ideal case, there is the presence of quantum corrections modifying the transport of momentum and the equation for the electric field. Section 5 studies the influence of the quantum terms on the equilibrium solutions. Exact solutions are found for translational invariance. Section 6 is devoted to the conclusions.

\section{Quantum Hydrodynamics in the Presence of Magnetic Fields}

For completeness, we begin with the derivation of the Wigner-Maxwell system providing a kinetic description for quantum plasmas in the presence of electromagnetic fields. 
For notational simplicity, we first consider a quantum hydrodynamics model for non zero magnetic fields in the case of a single species plasma. Extension to multi-species plasmas  is then straightforward. Our starting point is a statistical mixture with $N$ states described by the wave functions $\psi_\alpha = \psi_{\alpha}({\bf r},t)$, each with probability $p_\alpha$, with $\alpha = 1 \dots N$. Of course, $p_\alpha \geq 0$ and $\sum_{\alpha=1}^{N}p_\alpha = 1$. The wave functions obey the Schr\"odinger equation, 
\begin{equation}
\label{e1}
\frac{1}{2m}(-i\hbar\nabla - q{\bf A})^{2}\,\psi_\alpha + q\phi\,\psi_\alpha = i\hbar\frac{\partial\psi_\alpha}{\partial t} \,.
\end{equation}
Here we consider charge carriers of mass $m$ and charge $q$, subjected to possibly self-consistent scalar and vector potentials $\phi = \phi({\bf r},t)$ and ${\bf A} = {\bf A}({\bf r},t)$ respectively. For convenience in some calculations, we assume the Coulomb gauge, $\nabla\cdot{\bf A} = 0$.

From the statistical mixture, we construct the Wigner function $f = f({\bf r},{\bf p},t)$ defined as usual from 
\begin{equation}
\label{e2}
f({\bf r},{\bf p},t) = \frac{1}{(2\pi\hbar)^3}\sum_{\alpha=1}^{N}p_{\alpha}\int\,d{\bf s}\,\psi_{\alpha}^{*}({\bf r}+\frac{\bf s}{2})\,e^{\frac{i{\bf p}\cdot{\bf s}}{\hbar}}\,\psi_{\alpha}({\bf r}-\frac{\bf s}{2}) \,.
\end{equation}
After some long but simple calculations involving the Schr\"odinger equations for each $\psi_\alpha$ and the choice of the Coulomb gauge, we arrive at the following integro-differential equation for the Wigner function, 
\begin{eqnarray}
\label{e3}
&\strut& \frac{\partial f}{\partial t} + \frac{\bf p}{m}\cdot\nabla\,f = \\ &\strut& \frac{iq}{\hbar(2\pi\hbar)^3}\int\int d{\bf s}\,d{\bf p}'\,e^{\frac{i({\bf p}-{\bf p}')\cdot{\bf s}}{\hbar}}\,[\phi({\bf r}+\frac{\bf s}{2})-\phi({\bf r}-\frac{\bf s}{2})]\,f({\bf r},{\bf p}',t) + \nonumber 
\\
&\strut& \frac{iq^2}{2\hbar m(2\pi\hbar)^3}\int\int d{\bf s}\,d{\bf p}'\,e^{\frac{i({\bf p}-{\bf p}')\cdot{\bf s}}{\hbar}}\,[A^{2}({\bf r}+\frac{\bf s}{2})-A^{2}({\bf r}-\frac{\bf s}{2})]\,f({\bf r},{\bf p}',t) + \nonumber \\
&\strut& \frac{q}{2m(2\pi\hbar)^3}\,\,\nabla\cdot\int\int d{\bf s}\,d{\bf p}'\,e^{\frac{i({\bf p}-{\bf p}')\cdot{\bf s}}{\hbar}}\,[{\bf A}({\bf r}+\frac{\bf s}{2})-{\bf A}
({\bf r}-\frac{\bf s}{2})]\,f({\bf r},{\bf p}',t) \nonumber \\
&-& \frac{iq}{\hbar m(2\pi\hbar)^3}\,\,{\bf p}\cdot\int\int d{\bf s}\,d{\bf p}'\,e^{\frac{i({\bf p}-{\bf p}')\cdot{\bf s}}{\hbar}}\,[{\bf A}({\bf r}+\frac{\bf s}{2})-{\bf A}({\bf r}-\frac{\bf s}{2})]\,f({\bf r},{\bf p}',t)  \,. \nonumber
\end{eqnarray}
All macroscopic quantities like charge and current densities can be found taking appropriated moments of the Wigner function. This is analogous to classical kinetic theory, where charge and current densities are obtained from moments of the one-particle distribution function. Alternatively, we could have started from the complete many body wave function, defined a many body Wigner function and then obtained a quantum Bogoliubov-Born-Green-Kirkwood-Yvon hierarchy. With some closure hypothesis, in this way we arrive at a integro-differential equation for the one-particle Wigner function, which has to be supplemented by Maxwell equations. This is the Wigner-Maxwell system, which plays, in quantum physics, the same role the Vlasov-Maxwell system plays in classical physics. When the vector potential is zero, it reproduces the well known Wigner-Poisson system \cite{Drummond, Klimontovich}. In addition, in the formal classical limit when $\hbar \rightarrow 0$, the Wigner equation (\ref{e3}) goes to the Vlasov equation, 
\begin{equation}
\label{e4}
\frac{\partial f}{\partial t} + {\bf v}\cdot\nabla f + \frac{q}{m}({\bf E} + {\bf v}\times{\bf B})\cdot\frac{\partial f}{\partial{\bf v}} = 0 \,,
\end{equation}
where ${\bf v} = ({\bf p}-q{\bf A})/m$, ${\bf E} = - \nabla\phi - \partial{\bf A}/\partial t$ and ${\bf B} = \nabla\times{\bf A}$. However, notice that a initially positive definite Wigner function can evolve in such a way it becomes negative in some regions of phase space. Hence, it can not be considered as a true probability function. Nevertheless, all macroscopic quantities like charge, mass and current densities can be obtained from the Wigner function through appropriated moments. 

Equation (\ref{e3}) coupled to Maxwell equations provides a self-consistent kinetic description for a quantum plasma. As long as we know, it has been first obtained, with a different notation, in the work \cite{Arnold}. It has been rediscovered in \cite{Materdey}, in the case of homogeneous magnetic fields. Wigner functions appropriated to non zero magnetic fields have also been discussed, for instance, in \cite{Carruthers}-\cite{Bialynicki}, without the derivation of an evolution equation for the Wigner function alone. More recently, a different transport equation for Wigner functions appropriated to non zeros magnetic field and spin has been obtained in \cite{Saikin}. The starting point of this latter development, however, is the Pauli and not the Schr\"odinger equation as here and \cite{Arnold}. Finally, relativistic models for self-consistent charged particle systems with spin can be found in \cite{Masmoudi}. 

Most of the works dealing with quantum charged particle systems prefer to work with the wave functions and not directly with the Wigner function, as in \cite{Kumar}. The impressive form of (\ref{e3}) seems to support this approach. Indeed, probably (\ref{e3}) can be directly useful only in the linear or homogeneous magnetic field cases. This justifies the introduction of alternative descriptions. At the coast of the loss of some information about kinetic phenemena like Landau damping, we can simplify our model adopting a formal hydrodynamic formulation. Define the fluid density 
\begin{equation}
\label{e5}
n = \int\,d{\bf p}\,f \,,
\end{equation}
the fluid velocity
\begin{equation}
\label{e6}
{\bf u} = \frac{1}{mn}\int\,d{\bf p}\,({\bf p}-q{\bf A})\,f 
\end{equation}
and the pressure dyad 
\begin{equation}
\label{e7}
{\bf P} = \frac{1}{m^2}\int\,d{\bf p}\,({\bf p} - q{\bf A})\otimes({\bf p} - q{\bf A})\,f - n{\bf u}\otimes{\bf u} \,.
\end{equation}
We could proceed to higher order moments of the Wigner function, but (\ref{e5})-(\ref{e7}) are sufficient if we do not want to offer a detailed description of energy transport.

Taking the appropriated moments of the Wigner equation (\ref{e3}) and using the definitions (\ref{e5})-(\ref{e7}), we arrive at the following quantum hydrodynamic model, 
\begin{eqnarray}
\label{e8}
\frac{\partial n}{\partial t} &+& \nabla\cdot(n{\bf u}) = 0 \,, \\
\label{e9}
\frac{\partial{\bf u}}{\partial t} &+& {\bf u}\cdot\nabla{\bf u} =  - \frac{1}{n}\nabla\cdot{\bf P} + \frac{q}{m}({\bf E} + {\bf u}\times{\bf B}) \,.
\end{eqnarray}

Equations (\ref{e8})-(\ref{e9}) does not show in an obvious way any quantum effects, since $\hbar$ is not explicitly present there. To found the hidden quantum effects, we follow mainly the style of references \cite{Manfredi2}, \cite{Gasser1} and \cite{Lopez}, but now allowing for magnetic fields. In the definition (\ref{e2}) of the Wigner function, consider the decomposition 
\begin{equation}
\label{e10}
\psi_\alpha = \sqrt{n_\alpha}\,\,\,e^{iS_{\alpha}/\hbar} \,, 
\end{equation}
for real $n_\alpha = n_{\alpha}({\bf r},t)$ and $S_\alpha = S_{\alpha}({\bf r},t)$. Evaluating, the integral for the pressure dyad, we get a decomposition in terms of ``classical" ${\bf P}^C$ and ``quantum" ${\bf P}^Q$ contributions, 
\begin{equation}
\label{e11}
{\bf P} = {\bf P}^C + {\bf P}^Q \,,
\end{equation}
where 
\begin{eqnarray}
\label{a1}
{\bf P}^C &=& m\sum_{\alpha=1}^{N}{p_{\alpha}n_\alpha}({\bf u}_{\alpha} - {\bf u})\otimes({\bf u}_{\alpha} - {\bf u}) + \\ &+& 
m\sum_{\alpha=1}^{N}{p_{\alpha}n_\alpha}({\bf u}^{o}_{\alpha} - {\bf u}^{o})\otimes({\bf u}^{o}_{\alpha} - {\bf u}^{o}) \,, \nonumber \\
\label{a2}
{\bf P}^Q &=& - \frac{\hbar^{2}n}{4m}\nabla\otimes\nabla\,\ln\,n \,.
\end{eqnarray}
In the definitions of classical pressure dyad ${\bf P}^C$, we considered the kinetic fluid velocity associated to the wave function $\psi_\alpha$, 
\begin{equation}
\label{e15}
{\bf u}_{\alpha} = \frac{\nabla S_\alpha}{m} \,, 
\end{equation}
and the kinetic fluid velocity associated to the statistical mixture, 
\begin{equation}
\label{e16}
{\bf u} = \sum_{\alpha=1}^{N}\,\frac{p_{\alpha}n_\alpha}{n}{\bf u}_{\alpha} \,.
\end{equation}
In a similar way, the second term at the right hand side of equation (\ref{a1}) is constructed in terms of ${\bf u}_{\alpha}^o$, the osmotic fluid velocity associated to the wave function $\psi_\alpha$, 
\begin{equation}
\label{e17}
{\bf u}_{\alpha}^o = \frac{\hbar}{2m}\frac{\nabla n_\alpha}{n_\alpha} \,, 
\end{equation}
and ${\bf u}^o$, the osmotic fluid velocity associated to the statistical mixture, 
\begin{equation}
\label{e18}
{\bf u}^o = \sum_{\alpha=1}^{N}\,\frac{p_{\alpha}n_\alpha}{n}{\bf u}_{\alpha}^o \,.
\end{equation}
We also observe that in terms of the fluid density $n_\alpha$ of the state $\alpha$ the density $n$ of the statistical mixture is given by 
\begin{equation}
\label{e19}
n = \sum_{\alpha=1}^{N}\,p_{\alpha}n_\alpha \,.
\end{equation}
Notice that ${\bf P}^C$ a faithful classical pressure dyad, since it comes from dispersion of the velocities, vanishing for a pure state. Indeed, the classical pressure dyad is the sum of a kinetic part, arising from the dispersion of the kinetic velocities, and a osmotic part, arising from the dispersion of the osmotic velocities. However, ${\bf P}^C$ is not strictly classical, since it contains $\hbar$ through the osmotic velocities. In a sense, however, it is ``classical", since it comes from statistical dispersion of the velocities. 

In most cases, it suffices to take some equation of state for ${\bf P}^C$. For simplicity, from now on we assume a diagonal, isotropic form $P_{ij} = \delta_{ij}P$, where $P = P(n)$ is a suitable equation of state. Certainly, strong magnetic fields have to be treated more carefully, since they are associated to anisotropic pressure dyads. However, since we are mainly interested on the role of the quantum effects, we disregard such possibility here. 

Now inserting the preceding results for the pressure dyad into the momentum transport equation (\ref{e9}), we obtain the suggestive equation 
\begin{equation}
\label{e20}
\frac{\partial{\bf u}}{\partial t} + {\bf u}\cdot\nabla{\bf u} =  - \frac{1}{mn}\nabla P + \frac{q}{m}({\bf E} + {\bf u}\times{\bf B}) + \frac{\hbar^2}{2m^2}\nabla\left(\frac{\nabla^{2}\sqrt{n}}{\sqrt{n}}\right) \,.
\end{equation}
The equation of continuity (\ref{e8}) and the force equation (\ref{e20}) constitute our quantum hydrodynamic model for magnetized systems. All the quantum effects are contained in the last term of the equation (\ref{e20}), the so called Bohm potential. In comparison with standard fluid models for charged particle systems, the Bohm potential is the only quantum contribution, and the rest of the paper is devoted to study its consequences for magnetohydrodynamics. 

\section{Quantum Magnetohydrodynamics Model}

The equations from the last Section were written for a single species charged particle system. Now we generalize to a two species system. Consider electrons with fluid density $n_e$, fluid velocity ${\bf u}_e$, charge $-e$, mass $m_e$ and pressure $P_e$. In an analogous fashion, consider ions with fluid density $n_i$, fluid velocity ${\bf u}_i$, charge $e$, mass $m_i$ and pressure $P_i$. Proceeding as before, now starting from the Wigner equations for electrons and ions, we get the following bipolar quantum fluid model, 
\begin{eqnarray}
\label{e21}
\frac{\partial n_e}{\partial t} + \nabla\cdot(n_{e}{\bf u}_e) &=& 0 \,, \\
\label{e22}
\frac{\partial n_i}{\partial t} + \nabla\cdot(n_{i}{\bf u}_i) &=& 0 \,, \\
\label{e23}
\frac{\partial{\bf u}_e}{\partial t} + {\bf u}_{e}\cdot\nabla{\bf u}_e &=&  - \frac{\nabla P_{e}}{m_{e}n_{e}} - \frac{e}{m_e}({\bf E} + {\bf u}_{e}\times{\bf B}) + \nonumber \\ &+& \frac{\hbar^2}{2m^{2}_{e}}\nabla\left(\frac{\nabla^{2}\sqrt{n_e}}{\sqrt{n_e}}\right) - \nu_{ei}({\bf u}_e - {\bf u}_{i}) \,,\\
\label{e24}
\frac{\partial{\bf u}_i}{\partial t} + {\bf u}_{i}\cdot\nabla{\bf u}_i &=&  - \frac{\nabla P_{i}}{m_{i}n_{i}} + \frac{e}{m_i}({\bf E} + {\bf u}_{i}\times{\bf B}) + \nonumber \\ &+& \frac{\hbar^2}{2m^{2}_{i}}\nabla\left(\frac{\nabla^{2}\sqrt{n_i}}{\sqrt{n_i}}\right) - \nu_{ie}({\bf u}_i - {\bf u}_{e})\,.
\end{eqnarray}
In the equations (\ref{e23}-\ref{e24}), we have added some often used phenomenological terms to take into account for the momentum transport by collisions. The coefficients $\nu_{ei}$ and $\nu_{ie}$ are called collision frequencies for momentum transfer between electrons and ions \cite{Nicholson, Bittencourt}. For quasineutral plasmas, global momentum conservation in collisions imply $m_{e}\nu_{ei} = m_{i}\nu_{ie}$, so that $\nu_{ie} \ll \nu_{ei}$ when the ions are much more massive than electrons \cite{Nicholson, Bittencourt}.

Equations (\ref{e21})-(\ref{e24}) have to be supplemented by Maxwell equations, 
\begin{eqnarray}
\label{e25}
\nabla\cdot{\bf E} &=& \frac{\rho}{\varepsilon_0} \,,\\
\label{e26}
\nabla\cdot{\bf B} &=& 0 \,,\\
\label{e27}
\nabla\times{\bf E} &=& - \frac{\partial\bf B}{\partial t} \,,\\
\label{e28}
\nabla\times{\bf B} &=& \mu_{0}{\bf J} + \mu_{0}\varepsilon_{0}\frac{\partial\bf E}{\partial t} \,,
\end{eqnarray}
where the charge and current densities are given respectively by
\begin{equation}
\label{e29}
\rho = e\,(n_i - n_{e}) \,, \quad {\bf J} = e\,(n_{i}{\bf u}_i - n_{e}{\bf u}_{e}) \,.
\end{equation}
Equations (\ref{e21}-\ref{e29}) constitute our complete quantum hydrodynamic model,  allowing for magnetic fields. When ${\bf B} \equiv 0$, it goes to the well known quantum hydrodynamic model for bipolar charged particle systems.

Several possibilities of study are open starting from (\ref{e21}-\ref{e29}). Here we are interested in obtaining equations analogous to the classical magnetohydrodynamic equations. In some places, for the sake of clarity and to point exactly for the new contributions of quantum nature, we repeat some well known steps in the derivation of classical magnetohydrodynamics. To proceed in this direction, define the global mass density
\begin{equation}
\label{e30}
\rho_m = m_{e}n_e + m_{i}n_i 
\end{equation}
and the global fluid velocity 
\begin{equation}
\label{e31}
{\bf U} = \frac{m_{e}n_{e}{\bf u}_{e} + m_{i}n_{i}{\bf u}_{i}}{m_{e}n_{e} + m_{i}n_{i}}\,.
\end{equation}
With these definitions and proceeding like in any plasma physics book \cite{Nicholson, Bittencourt}, we obtain the following equations for $\rho_m$ and ${\bf U}$, 
\begin{eqnarray}
\label{e32}
\frac{\partial\rho_m}{\partial t} + \nabla\cdot(\rho_{m}{\bf U}) &=& 0 \,,\\
\rho_{m}(\frac{\partial{\bf U}}{\partial t} + {\bf U}\cdot\nabla{\bf U}) &=& - \nabla\cdot{\bf\Pi} + 
\rho{\bf E} + {\bf J}\times{\bf B} + \nonumber \\
\label{e33}
&+& \frac{\hbar^{2}n_{e}}{2m_{e}}\nabla\left(\frac{\nabla^{2}\sqrt{n_{e}}}{\sqrt{n_{e}}}\right) + \frac{\hbar^{2}n_{i}}{2m_{i}}\nabla\left(\frac{\nabla^{2}\sqrt{n_{i}}}{\sqrt{n_{i}}}\right) \,,
\end{eqnarray}
for 
\begin{equation}
\label{e34}
{\bf\Pi} = P\,{\bf I} + \frac{m_{e}m_{i}n_{e}n_{i}}{\rho_m}({\bf u}_e - {\bf u}_{i})\otimes({\bf u}_e - {\bf u}_{i}) \,, 
\end{equation}
where $P = P_{e} + P_{i}$ and where ${\bf I}$ is the identity matrix. In equations (\ref{e33}-\ref{e34}), the electronic and ionic densities are defined in terms of the mass and charge densities according to 
\begin{equation}
\label{e35}
n_e = \frac{1}{m_i + m_{e}}\,\,(\rho_m - \frac{m_{i}}{e}\rho) \,, \quad n_i = \frac{1}{m_i + m_{e}}\,\,(\rho_m + \frac{m_{e}}{e}\rho) \,.
\end{equation}
We can simplify (\ref{e33}) considerably assuming, as usual, quasi-neutrality ($\rho = 0$ so that $n_e = n_i$), $P_e = P_i = P/2$ and neglecting $m_e$ in comparison to $m_i$ whenever possible. In addition, disregarding the last term at the right hand side of (\ref{e34}), we obtain 
\begin{equation}
\label{e36}
\frac{\partial{\bf U}}{\partial t} + {\bf U}\cdot\nabla{\bf U} = -\frac{1}{\rho_m}\nabla P + \frac{1}{\rho_m}{\bf J}\times{\bf B} + \frac{\hbar^{2}}{2m_{e}m_{i}}\nabla\left(\frac{\nabla^{2}\sqrt{\rho_m}}{\sqrt{\rho_m}}\right) \,.
\end{equation}
Under the same assumptions and following the standard derivation of magnetohydrodynamics \cite{Nicholson, Bittencourt}, we obtain the following equation for the current ${\bf J}$, 
\begin{equation}
\label{e37}
\frac{m_{e}m_{i}}{\rho_{m}e^2}\frac{\partial{\bf J}}{\partial t} - \frac{m_{i}\nabla P}{\rho_{m}e} = {\bf E} + {\bf U}\times{\bf B} - \frac{m_{i}}{\rho_{m}e}\,{\bf J}\times{\bf B} - \frac{\hbar^{2}}{2e m_{e}}\nabla\left(\frac{\nabla^{2}\sqrt{\rho_m}}{\sqrt{\rho_m}}\right) - \frac{1}{\sigma}\,{\bf J} \,,
\end{equation}
where $\sigma = \rho_{m}e^{2}/(m_{e}m_{i}\nu_{ei})$ is the longitudinal electrical conductivity. Equation (\ref{e37}) is the quantum version of the generalized Ohm's law \cite{Nicholson, Bittencourt}. The continuity equation (\ref{e32}), the force equation (\ref{e36}), the quantum version of the generalized Ohm's law (\ref{e37}), an equation of state for $P$, plus Maxwell equations, provides a full system of quantum magnetohydrodynamic equations. However, it is probably still complicated and in the next section we propose some approximations in the same spirit of those of classical magnetohydrodynamics.

\section{Simplified and Ideal Quantum Magnetohydrodynamic Equations}
Usually \cite{Nicholson, Bittencourt}, the left-hand side of the equation (\ref{e37}) is neglected in the cases of slowly varying processes and small pressures. Also, for slowly varying and high conductivity problems , the displacement current can be neglected in Amp\`ere's law. Finally, we assume an equation of state appropriated for adiabatic processes. This provides a complete system of simplified quantum magnetohydrodynamic equations, which we collect here for convenience, 
\begin{eqnarray}
\label{e38}
\frac{\partial\rho_m}{\partial t} &+& \nabla\cdot(\rho_{m}{\bf U}) = 0 \,,\\
\label{e39}
\frac{\partial{\bf U}}{\partial t} &+& {\bf U}\cdot\nabla{\bf U} = - \frac{1}{\rho_m}\nabla  P + \frac{1}{\rho_m}{\bf J}\times{\bf B} + \frac{\hbar^{2}}{2m_{e}m_{i}}\nabla(\frac{\nabla^{2}\sqrt{\rho_m}}{\sqrt{\rho_m}}) \,,\\
\label{e40}
\nabla P &=& V_{s}^{2}\nabla\rho_m \,,\\
\label{e41}
\nabla&\times&{\bf E} = - \frac{\partial{\bf B}}{\partial t} \,,\\
\label{e42}
\nabla&\times&{\bf B} = \mu_{0}{\bf J} \,,\\
\label{e43}
{\bf J} &=& \sigma[{\bf E} + {\bf U}\times{\bf B} - \frac{m_{i}}{\rho_{m}e}\,{\bf J}\times{\bf B} - \frac{\hbar^{2}}{2e m_{e}}\nabla(\frac{\nabla^{2}\sqrt{\rho_m}}{\sqrt{\rho_m}})] \,.
\end{eqnarray}
In equation (\ref{e40}), $V_s$ is the adiabatic speed of sound of the fluid. Gauss law can be regarded as the initial condition for Faraday's law. Also notice that the Hall term ${\bf J}\times{\bf B}$ at (\ref{e43}) is often neglected in magnetohydrodynamics. 

Inserting (\ref{e40}) into (\ref{e39}), we are left with a system of 13 equations for 13 unknowns, namely, $\rho_m$ and the components of ${\bf U}, {\bf J}, {\bf B}$ and ${\bf E}$. 
This is our quantum magnetohydrodynamics model. In comparison to classical  magnetohydrodynamics, the difference of the present model rests on the presence of two quantum corrections, the last terms at equations (\ref{e39}) and (\ref{e43}). 

In the ideal magnetohydrodynamics approximation, we assume an infinite conductivity and neglect the Hall force at (\ref{e43}). This provides the following ideal quantum  magnetohydrodynamics model, 
\begin{eqnarray}
\label{e44}
{\bf E} = - {\bf U}\times{\bf B} &+& \frac{\hbar^{2}}{2e m_{e}}\nabla(\frac{\nabla^{2}\sqrt{\rho_m}}{\sqrt{\rho_m}}) \,, \\
\rho_{m}(\frac{\partial{\bf U}}{\partial t} + {\bf U}\cdot\nabla{\bf U}) &=& - \nabla P + \frac{1}{\mu_0}(\nabla\times{\bf B})\times{\bf B} + \nonumber \\ \label{e45} &+& \frac{\hbar^{2}\rho_m}{2m_{e}m_{i}}\nabla(\frac{\nabla^{2}\sqrt{\rho_m}}{\sqrt{\rho_m}}) \,,\\
\label{e46}
\frac{\partial{\bf B}}{\partial t} &=& \nabla\times({\bf U}\times{\bf B}) \,,
\end{eqnarray}
supplemented by the continuity equation (\ref{e38}) and the equation of state (\ref{e40}).  

Taking into account (\ref{e40}), equations (\ref{e45}-\ref{e46}) plus the continuity equation provides a system of 7 equations for 7 unknowns, namely, $\rho_m$ and the components of ${\bf U}$ and ${\bf B}$. This is our ideal quantum magnetohydrodynamics model. In comparison to classical ideal magnetohydrodynamics, the difference of the present model rests on the presence of a quantum correction, the last term at equation (\ref{e45}). Interestingly, taking the curl of (\ref{e44}) makes disappear one of the quantum correction terms present in the non ideal quantum magnetohydrodynamics. This leads to a dynamo equation (\ref{e46}) identical to that of classical magnetohydrodynamics. Consequently, for infinite conductivity the magnetic field lines are still frozen to the fluid, even allowing for the quantum corrections proposed here. In fact, even for finite conductivity, the diffusion of magnetic field lines is described by the same diffusion equation as that of classical magnetohydrodynamics.  This comes from the fact that the quantum correction disappear after taking the curl of both sides of (\ref{e43}), neglecting the Hall term and assuming a constant $\sigma$ as usual. However, a further quantum correction on the electric field still survives through (\ref{e44}). 

In order to obtain a deeper understanding of the importance of quantum effects, we propose the following rescaling for our ideal quantum magnetohydrodynamic equations, 
\begin{eqnarray}
\bar{\rho}_m &=& \rho_{m}/\rho_0 \,, \quad \bar{\bf U} = {\bf U}/V_A \,, \quad \bar{\bf B} = {\bf B}/B_0 \,, \nonumber \\
\label{e48}
\bar{\bf r} &=& \Omega_{i}{\bf r}/V_A \,, \quad \bar{t} = \Omega_{i}t \,,
\end{eqnarray}
where $\rho_0$ and $B_0$ are the equilibrium mass density and magnetic field. In addition, $V_A = (B_{0}^{2}/(\mu_{0}\rho_{0}))^{1/2}$ is the Alfv\'en velocity and $\Omega_i = eB_{0}/m_i$ is the ion cyclotron velocity. We justify the chosen rescaling in the following way. In magnetohydrodynamics, the Alf\'en velocity provides a natural velocity scale. Also, since we deal with low frequency problems, $\Omega_{i}^{-1}$ is a reasonable  candidate for a natural time scale. These velocity and time scales induces the length scale $V_{A}/\Omega_{i}$, as shown in (\ref{e48}). 

Applying the rescaling (\ref{e48}) to the ideal quantum magnetohydrodynamic equations, we obtain the following non dimensional model, 
\begin{eqnarray}
\label{e49}
\frac{\partial\bar{\rho}_m}{\partial t} &+& \nabla\cdot(\bar{\rho}_{m}\bar{\bf U}) = 0 \,,\\
\bar{\rho}_{m}(\frac{\partial\bar{\bf U}}{\partial t} + \bar{\bf U}\cdot\nabla\bar{\bf U}) &=& - \frac{V_{s}^2}{V_{A}^2}\nabla\bar{\rho}_m + (\nabla\times\bar{\bf B})\times\bar{\bf B} + \nonumber \\ \label{e50} &+& \frac{H^{2}\bar{\rho}_m}{2}\nabla(\frac{\nabla^{2}\sqrt{\bar{\rho}_m}}{\sqrt{\bar{\rho}_m}}) \,,\\
\label{e51}
\frac{\partial\bar{\bf B}}{\partial t} &=& \nabla\times(\bar{\bf U}\times\bar{\bf B}) \,,
\end{eqnarray}
where 
\begin{equation}
\label{e52}
H = \frac{\hbar\Omega_i}{\sqrt{m_{e}m_{i}}\,\,V_{A}^{2}} 
\end{equation}
is a non dimensional parameter measuring the relevance of quantum effects. Numerically, using M.K.S. units, we have $H = 3.42 \times 10^{-30} \,\,n_{0}/B_{0}$, where $n_0$ is the ambient particle density. While for ordinary plasmas $H$ is negligible, for dense astrophysical plasmas \cite{Chabrier}-\cite{Jung}, with $n_0$ about $10^{29} - 10^{34}\,\, m^{-3}$, $H$ can be of order unity or more. Hence, in dense astrophysical plasmas like the atmosphere of neutron stars or the interior of massive white dwarfs, quantum corrections to magnetohydrodynamics can be of experimental importance. Similar comments apply to our non ideal quantum magnetohydrodynamics model. However, even for moderate $H$ quantum effects can be negligible if the density is slowly varying in comparison with some typical length scale, due to the presence of a third order derivative at the Bohm potential. This is in the same spirit of the Thomas-Fermi approximation. 

\section{Quan\-tum Ideal Magnetostatic E\-qui\-li\-brium}

There is a myriad of developments based on classical magnetohydrodynamics (linear and nonlinear waves, dynamo theory and so on) and we shall not attempt to reproduce all the quantum counterparts of these subjects in the framework of our model. We will be restricted to just one subject, namely the construction of exact equilibria for ideal quantum magnetohydrodynamics, with no attempt to study the important question of the stability of the equilibria. 

Assuming that ${\bf U} = 0$ and that all quantities are time-independent, the ideal quantum magnetohydrodynamic equations (\ref{e44}-\ref{e46}) becomes 
\begin{eqnarray}
\label{e53}
{\bf E} &=& \frac{\hbar^{2}}{2e m_{e}}\nabla(\frac{\nabla^{2}\sqrt{\rho_m}}{\sqrt{\rho_m}}) \,, \\
\label{e54}
\nabla P &=& \frac{1}{\mu_0}(\nabla\times{\bf B})\times{\bf B} + 
\frac{\hbar^{2}\rho_m}{2m_{e}m_{i}}\nabla(\frac{\nabla^{2}\sqrt{\rho_m}}{\sqrt{\rho_m}}) \,.
\end{eqnarray}
According to (\ref{e53}), the equilibrium solutions of ideal quantum magnetohydrodynamics are not electric field free any longer. In addition, equation (\ref{e54}) has an quantum correction that invalidate the classical magnetic surface equation for ${\bf B}\cdot\nabla{\bf B} = 0$, namely $P + B^{2}/(2\mu_{0}) =$ cte. 

Equation (\ref{e54}) together with an equation of state is the key for the search of equilibrium solutions. We will try to follow, as long as possible, the strategy of reference \cite{Hamabata} for classical magnetostatic equilibria. Inspired by well known classical solutions \cite{Hamabata}, assume a translationally invariant solution of the form 
\begin{eqnarray}
\label{e55}
P &=& P(r,\varphi) \,, \quad \rho_m = \rho_{m}(r,\varphi) \,, \\
\label{e56}
{\bf B} &=& \nabla A(r,\varphi)\times\hat{z} + B_{z}(r,\varphi)\hat{z} \,,
\end{eqnarray}
using cylindrical coordinates and where $A = A(r,\varphi)$ and $B_{z} = B_{z}(r,\varphi)$ as well as the pressure and the mass density are functions of $(r,\varphi)$ only. 

Substituting the proposal (\ref{e55}-\ref{e56}) into (\ref{e54}), we get, for the radial and azimuthal components of this equation, 
\begin{equation}
\label{e57}
\nabla(P + \frac{B_{z}^2}{2\mu_0}) = - \frac{1}{\mu_0}\,\nabla A\,\,\nabla^{2}A + \frac{\hbar^{2}\rho_{m}}{2m_{e}m_{i}}\,\,\nabla(\frac{\nabla^{2}\sqrt{\rho_m}}{\sqrt{\rho_m}}) \,,
\end{equation}
while, for the $z$ component, the result is 
\begin{equation}
\label{e58}
\frac{\partial(B_{z},A)}{\partial(r,\varphi)} = 0 \,.
\end{equation}
In (\ref{e58}) and in what follows, we used the definition of Jacobian,
\begin{equation}
\label{e59}
\frac{\partial(B_{z},A)}{\partial(r,\varphi)} = 
\left(\matrix{\frac{\partial\,B_z}{\partial r} & \frac{\partial\,B_z}{\partial \varphi}\cr 
\frac{\partial\,A}{\partial r} & \frac{\partial\,A}{\partial \varphi}\cr}\right) \,.
\end{equation}
From (\ref{e58}), we obtain 
\begin{equation}
\label{e60}
B_z = B_{z}(A) \,. 
\end{equation}
Taking into account (\ref{e57}) and the fact that $B_z$ is a function of $A$, it follows that 
\begin{equation}
\label{e61}
\frac{\partial(P,A)}{\partial(r,\varphi)} = \frac{\hbar^{2}\rho_{m}}{2m_{e}m_{i}}\frac{\partial(\nabla^{2}\sqrt{\rho_{m}}\,\,/\sqrt{\rho_m}\,\,
,A)}{\partial(r,\varphi)} \,.
\end{equation}
In the classical limit $\hbar \rightarrow 0$, the right hand of (\ref{e61}) vanishes, implying just the functional relationship $P = P(A)$. In the present work, we still postulate 
\begin{equation}
\label{e62}
P = P(A) \,,
\end{equation}
so that, from (\ref{e61}), we have 
\begin{equation}
\label{e63}
\frac{\nabla^{2}\sqrt{\rho_{m}}}{\sqrt{\rho_{m}}} = F(A) \,,
\end{equation}
where $F = F(A)$ is an arbitrary function. 

The last equation is a distinctive feature of ideal quantum magnetohydrodynamic equilibrium. Indeed, (\ref{e63}) would not be necessary if $\hbar = 0$ in (\ref{e61}). Hence, even if $\hbar$ is not present in (\ref{e63}), this equation has a quantum nature, with important implications in what follows. The reason why $\hbar$ does not appear in (\ref{e63}) is that it factor at the right hand side of (\ref{e61}). 

From (\ref{e62}) and some subjacent equation of state, $P = P(\rho_{m})$, we deduce 
\begin{equation}
\label{e64}
\sqrt{\rho_{m}} = G(A) \,,
\end{equation}
for some function $G = G(A)$. Plugging this into (\ref{e63}), the result is 
\begin{equation}
\label{e65}
\frac{G'}{G}\,\nabla^{2}A + \frac{G''}{G}\,(\nabla\,A)^2 = F(A) \,, 
\end{equation}
where the prime denotes derivation with respect to $A$. 

Coming back to (\ref{e57}), we obtain 
\begin{equation}
\label{e66}
\nabla^{2}A = \mu_{0}[- K'(A) + \frac{\hbar^{2}}{2m_{e}m_{i}}\,\,G^{2}F'(A)] \,,
\end{equation}
where we have defined
\begin{equation}
\label{e67}
K = K(A) = P(A) + \frac{B_{z}^{2}(A)}{2\mu_0} \,.
\end{equation}

Recapitulating, we have three four functions of $A$ to be stipulated, namely $F$, $G$, $K$ and $P$. However, $A$ satisfy two different equations, (\ref{e65}) and (\ref{e66}). Once $A$ is found, all other quantities (pressure, mass density, electromagnetic field) comes as consequences. 

A reasonable choice is to take $G$ as a linear function of $A$, since then (\ref{e65}) becomes linear in the derivatives. Hence, let
\begin{equation}
\label{e68}
G = k_{1}A + k_2 \,, \quad k_1 \neq 0 \,,
\end{equation}
for numerical constants $k_1$ and $k_2$. We take $k_1 \neq 0$ since $k_1 = 0$ would imply $F = 0$, making disappear the quantum correction at (\ref{e66}). With the choice (\ref{e68}), the couple (\ref{e65}-\ref{e66}) becomes 
\begin{eqnarray}
\label{e69}
\nabla^{2} A &=& \frac{1}{k_1}\,(k_{1}A + k_{2})\,F(A) \,, \\
\label{e70}
\nabla^{2} A &=& \mu_{0}\,[-K'(A) + \frac{\hbar^{2}}{2m_{e}m_{i}}\,(k_{1}A+k_{2})^{2}\,F'(A)] \,.
\end{eqnarray}
The right hand sides of (\ref{e69}) and (\ref{e70}) should coincide, implying
\begin{equation}
\label{e71}
K'(A) = \frac{\hbar^{2}}{2m_{e}m_{i}}\,(k_{1}A + k_{2})^{2}\,F'(A) - \frac{1}{\mu_{0}k_{1}}\,(k_{1}A+k_{2})\,F(A) \,.
\end{equation}
The last equation define $K$ up to an unimportant numerical constant. 

Equation (\ref{e69}) is the key equation for our translationally invariant magnetostatic equilibria. For a given $F(A)$ and solving (\ref{e69}) for $A$, all other quantities follows for a known equation of state. Indeed, knowing $A$ we can construct the radial and azimuthal components of the magnetic field through (\ref{e56}) and the mass density from (\ref{e64}). From the mass density and the equation of state, we obtain the pressure $P$. Proceeding, equation (\ref{e71}) yields $K(A)$ and then the $z$ component of the magnetic field through (\ref{e67}). Finally, the electric field follows from (\ref{e53}) and the current density from the curl of the magnetic field. The free ingredients to be chosen to construct explicitly the exact solution are the function $F(A)$ and the equation of state, and the numerical constants $k_1$ and $k_2$. Other possibilities can be explored if we do not restrict to linear $G(A)$ functions as in (\ref{e68}), but then $A$ will not satisfy an linear in the derivatives equation. 

\subsection{An Explicit Exact Solution}
An interesting case of explicit solution is provided by the choice
\begin{equation}
\label{e72}
F(A) = \frac{k_{1}\,B_{0}\,(1-\varepsilon^{2}k)}{k_{1}A + k_{2}}\,\,e^{-2kA/B_{0}} \,,
\end{equation}
where $B_0$ is an arbitrary constant magnetic field, $k$ is an arbitrary constant with dimensions of an inverse length and $0 \leq \varepsilon < 1$. With the choice (\ref{e72}), the equation  (\ref{e69}) traduces into the Liouville equation, 
\begin{equation}
\label{e73}
\nabla^{2}A = (1-\varepsilon^{2})\,B_{0}\,k\,e^{-2kA/B_{0}} \,,
\end{equation}
which admits the exact cat eye solution
\begin{equation}
\label{e74}
A = \frac{B_{0}}{k}\,\,\ln[\cosh(kr\,cos\varphi) + \varepsilon\,\cos(kr\sin\varphi)] \,.
\end{equation}

All other relevant quantities follows from this exact solution following the recipe just stated. The mass density, from (\ref{e64}), is
\begin{equation}
\label{e75}
\rho_m = [\frac{k_{1}B_{0}}{k}\,\,\ln(\cosh(kr\,\cos\varphi) + \varepsilon\,\cos(kr\sin\varphi)) + k_{2}]^2 \,,
\end{equation}
while the radial and azimuthal components of the magnetic field follows from (\ref{e56}), 
\begin{eqnarray}
\label{e76}
B_r &=& - \frac{B_{0}\,[\sin\varphi\,\sinh(kr\,\cos\varphi) +  \varepsilon\cos\varphi\sin(kr\,\sin\varphi)]}{[\cosh(kr\,\cos\varphi) + \varepsilon\cos(kr\,\sin\varphi)]} \,, \\
\label{e77}
B_\varphi &=& 
- \frac{B_{0}\,[\cos\varphi\,\sinh(kr\,\cos\varphi) -  \varepsilon\sin\varphi\sin(kr\,\sin\varphi)]}{[\cosh(kr\,\cos\varphi) + \varepsilon\cos(kr\,\sin\varphi)]} \,.
\end{eqnarray}
Assuming an adiabatic equation of state, $P = V_{s}\rho_m$, we get, from (\ref{e67}), 
\begin{eqnarray}
\label{e78}
B_{z}^2 &=& B_{0}^2 - 2\mu_{0}\,V_{s}^{2}\,(k_{1}A+k_{2})^{2} + \\ &+&(1 - \varepsilon^{2})\,k^{2}\,e^{-2A}\,\,[1 + \mu_{0}k_{1}\,(k_{1}+k_{2})\hbar^{2}/m + \mu_{0}\,\hbar^{2}\,k_{1}^{2}\,A/m] \,,
\end{eqnarray}
with $A$ given by the cat eye solution (\ref{e74}). If desired, the electric field and the current density can then be calculated via (\ref{e53}) and Amp\`ere's law respectively. 
In figure 1, we show the contour plot of the function $A$ given by (\ref{e74}), while in figure 2 we show the corresponding mass density. The parameters chosen were $B_0 = 1$, $k = 1$, $\varepsilon = 0.9$, $k_1 = 1$ and $k_2 = 0$. These graphics shows coherent, periodic patterns resembling quantum periodic solutions arising in other quantum plasma systems \cite{Manfredi2}. Similar graphics can be easily obtained for the electromagnetic field and other macroscopic quantities derivable from the cat eye solution (\ref{e74}). 

\section{Conclusion}

In this work, we have obtained a quantum version of magnetohydrodynamics starting from a quantum hydrodynamics model with nonzero magnetic fields. In view of its simplicity, this magnetic quantum hydrodynamics model seems to be an attractive alternative to the Wigner magnetic equation of Section 2. The infinite conductivity approximation leads to an ideal quantum magnetohydrodynamics. For very dense plasmas and not to strong magnetic fields, the quantum corrections to magnetohydrodynamics can be relevant, as apparent from the parameter $H$ derived in Section 4. Under a number of suitable assumptions, we have derived some exact translationally invariant quantum ideal magnetostatic solutions. More general quantum ideal magnetostatic equilibria can be conjectured, in particular for axially symmetric situations. In addition, we have left a full investigation of linear waves to future works.

\vskip 1cm
\noindent{\bf Acknowledgments}\\
We thanks the Brazilian agency Conselho Nacional de
Desenvolvimento Cien\-t\'{\i}\-fi\-co e Tecn\'ologico (CNPq) for
financial support.

\newpage

\begin{figure}
\centering \resizebox{4in}{!}{\includegraphics{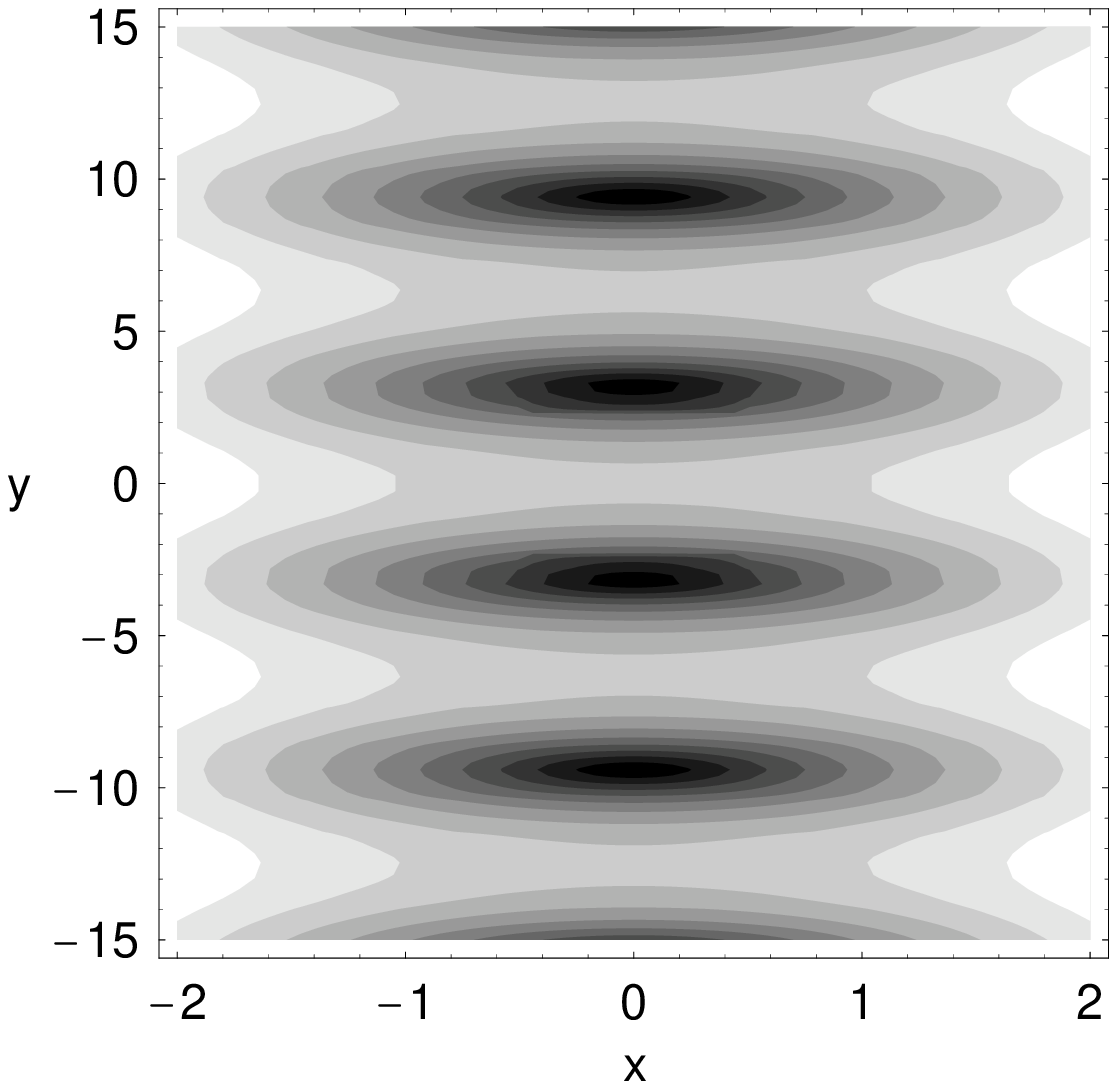}}
\caption{
Contour plot of the cat eye solution $A$ given by (\ref{e74}). 
The parameters are $B_0 = 1$, $k = 1$ and $\varepsilon = 0.9$.}
\label{fig1}
\end{figure}
\begin{figure}
\centering \resizebox{4in}{!}{\includegraphics{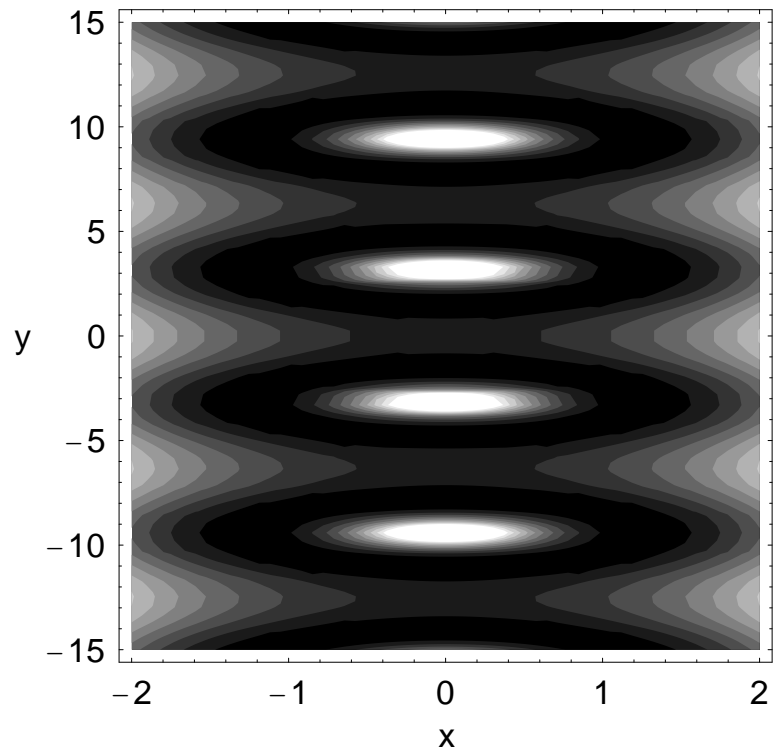}}
\caption{
Contour plot of the mass density $\rho_m$ given by (\ref{e75}). 
The parameters are $B_0 = 1$, $k = 1$, $\varepsilon = 0.9$, $k_1 = 1$ and $k_2 = 0$.}
\label{fig2}
\end{figure}
\end{document}